\documentclass[twocolumn,aps,prl,preprintnumbers,nofootinbib]{revtex4}
\usepackage{axodraw}
\usepackage{stmaryrd}
\usepackage{graphicx}
\usepackage{amsmath}
\usepackage{amssymb}
\newcommand{\nbb}[0]{{0\nu}\beta\beta}

\newcommand{\lbb}[0]{\Lambda_{\beta\beta}}
\newcommand{\mw}[0]{M_{\text{W}}}
\newcommand{\mr}[0]{M_{\text{R}}}
\newcommand{\nr}[0]{N_{\text{R}}}

\newcommand{\mn}[0]{m_\nu}
\newcommand{\lh}[0]{\Lambda_{\text{H}}}
\newcommand{\lr}[0]{\Lambda_{\text{R}}}
\newcommand{\wrt}[0]{W_{\text{R}}}
\newcommand{\wl}[0]{W_{\text{L}}}
\newcommand{\ls}[0]{\Lambda_{\text{S}}}
\newcommand{\gf}[0]{G_{\text{F}}}
\newcommand{\ga}[0]{g_{\text{A}}}
\newcommand{\gv}[0]{g_{\text{V}}}

\newcommand{\fp}[0]{f_\pi}

\newcommand{\Eq}[1]{Eq.~(\ref{#1})}

\newcommand{\Fig}[1]{Fig.~\ref{#1}}

\newcommand{\tl}[0]{\text{L}}

\begin{document}

%%{\noindent
%%\hfill
%%\begin{flushright}
%%CALT-68-2372\\
%%hep-ph/0409235\\
%%\end{flushright}
%%}

\author{Gary Pr{\'e}zeau}
\affiliation{Jet Propulsion
  Laboratory/California Institute of  Technology, 4800 Oak Grove Dr,
  Pasadena, CA 91109, USA}

\title{Light neutrino and heavy particle exchange in $\nbb$-decay}

%\date

\begin{abstract}

A simple and precise method is presented to compare contributions
to neutrinoless double-beta decay ($\nbb$-decay) from heavy particle
exchange and light Majorana neutrino exchange.  This procedure makes
no assumptions about the momentum transfer between the two nucleons
involved in the $\nbb$-decay process.  It is shown that for a
general particle physics model, the characteristic $\nbb$-decay scale
$> 4.4$~TeV when all the coupling constants are assumed to be natural
and of ${\cal{O}}(1)$.

\end{abstract}

%\narrowtext
\maketitle

%\arabic{figure}
\pagenumbering{arabic}

\begin{figure}[tb]
\resizebox{16cm}{!}{\includegraphics*[38,455][578,680]{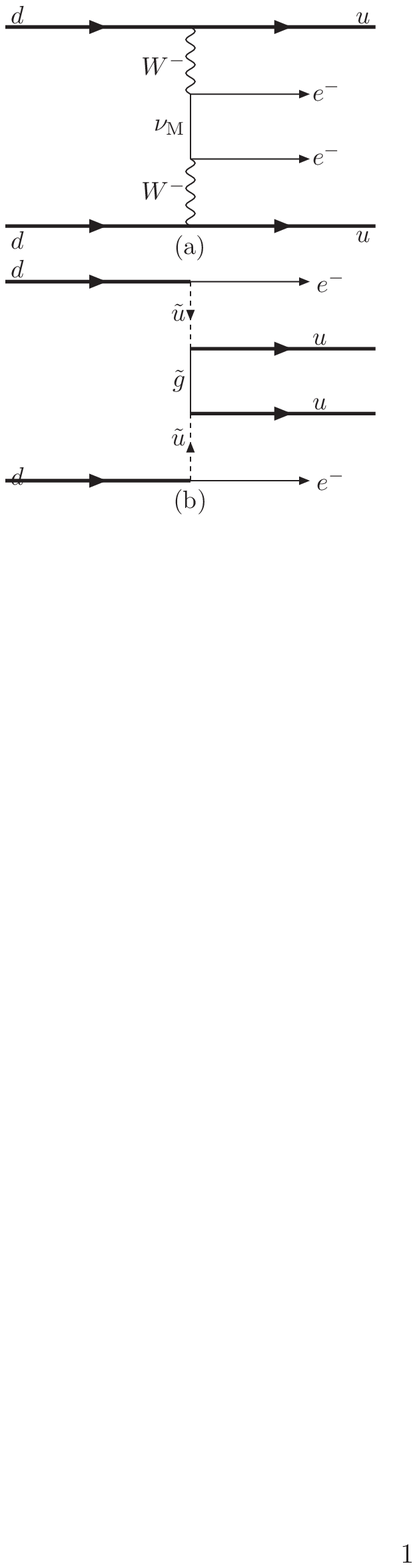}}
\caption{ \sl \footnotesize a) $\nbb$-decay through the
    exchange of a light Majorana neutrino.  b)
  $\nbb$-decay through the exchange of heavy particles, in this case two
  squarks and a gluino in RPV SUSY.}\label{intrographs}
\end{figure}

With the discovery of neutrino oscillations a few years
ago~\cite{Fukuda:1998mi,Ahmad:2002jz,Eguchi:2002dm}, the 
fundamental question of whether at least some neutrinos have
mass has been answered in the positive.  The parallel questions of
a) the magnitudes of the neutrino masses and b) the nature of the
neutrino mass matrix remain to be answered.  If the neutrinos have a
Dirac mass matrix, then lepton number is not violated by neutrino
interactions while the right-handed neutrinos and left-handed
anti-neutrinos are electroweak singlets.
Alternatively, for a Majorana neutrino mass matrix, lepton number is
violated by two units, and processes like 
neutrinoless double-beta decay ($\nbb$-decay) are permitted as
demonstrated by the Feynman diagram of \Fig{intrographs}a.

The observation of $\nbb$-decay would shed light on the neutrino mass
magnitude and whether it is Dirac or Majorana, but additional input
would be required.  Indeed, a number of particle physics model beyond
the standard model (SM) have lepton-number violating (LNV) operators
that allow $\nbb$-decay through the exchange of  heavy particles such
as the neutralino 
\cite{Mohapatra:su,Vergados:1986td,Hirsch:1995ek} or a heavy
right-handed neutrino 
\cite{Vergados:pv,Mohapatra:1974gc,Senjanovic:1975rk}.  Thus, the
observation of $\nbb$-decay would provide a unique window on physics
beyond the SM with broad implications for LNV particle
physics models, and the way neutrino masses are generated within them.
It follows
that disentangling contributions to $\nbb$-decay due to light Majorana
neutrino exchange from the heavy particle contributions is crucial if
one is to use this data to constrain the neutrino mass matrix and the
models that generate it and LNV.

%%%%%%%%%%%%%%%%%%%%%%%%%%%%%%%%%%%%%%%%%%%%%%%%%%%%%%%%%%%%%%%%%%%%
%%\begin{center}
%%\end{center}
%%%%%%%%%%%%%%%%%%%%%%%%%%%%%%%%%%%%%%%%%%%%%%%%%%%%%%%%%%%%%%%%%%%%

Although it has been known for some time that heavy particle exchange
and light Majorana neutrino exchange contributions to $\nbb$-decay can
be comparable, the comparisons have usually been performed by making
assumptions about the momemtum flow through the light Majorana
neutrino and estimating orders of
magnitude~\cite{Mohapatra:1998ye,Cirigliano:2004tc}. 
In this current work, a simple and more precise procedure is presented
to compare the relative importance of both processes to $\nbb$-decay. 
Operators stemming from light neutrino 
exchange that have precisely the same form as the leading order (LO)
heavy particle exchange $\nbb$-decay operators~\cite{Prezeau:2003xn}
are derived and used 
for the comparison.  Hence, there is no need to make an estimate of
the average momentum flowing through $\nu_{\text{M}}$ in
\Fig{intrographs}a.

\begin{figure}[tb]
\resizebox{17 cm}{!}{\includegraphics*[60,340][590,685]{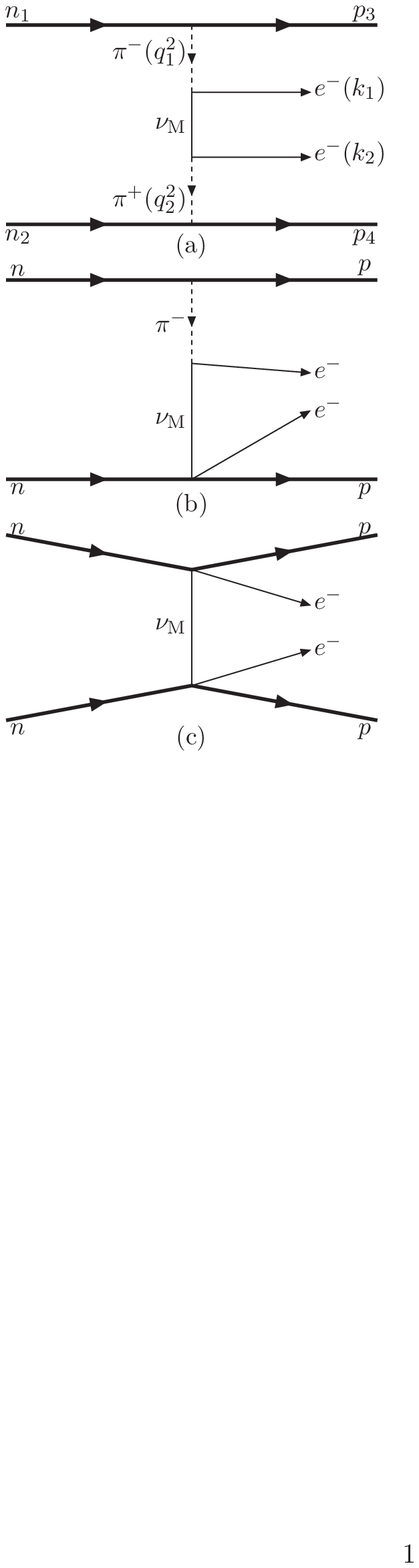}}
\caption{$\nbb$-decay operators with light Majorana neutrino
  exchange.  There exists another diagram like (b) where the pion and
  neutrino lines are exchanged.} \label{pionex}
\end{figure}
This might seem counter-intuitive since
 $\nbb$-decay  mediated by light neutrino exchange is suppressed by
 the ratio of the neutrino mass to $Q^2$ (the momentum 
squared flowing through the neutrino
 propagator~$|Q|\sim$~100~MeV).\footnote{Of course, it is not the
 neutrino mass that appears, but $m_{\beta\beta}$, the sum over the
 neutrino mass eigenstates multiplied by phases that may generate further
 suppression when squared.}
 In contrast, $\nbb$-decay is suppressed by $\lbb$ when it occurs 
through heavy particle exchange, where $\lbb$ is the heavy scale
(typically of the order of 1~TeV or larger) that
characterizes the strength of the $\nbb$-decay operator.  

The SM low-energy effective Lagrangian with Majorana neutrinos
%%%%%%%%%%%%%%%%%%%%%%%%%%%%%%%%%%%%%%%%%%%%%%%%%%%%%%%%%%%%%%%%%%%%
\begin{eqnarray}\label{SMLag}
{\cal{L}}_{\text{SM}} \doteq 
4\frac{\gf}{\sqrt{2}}\bar{u}_{\text{L}}\gamma^\mu d_\tl ~
\bar{e}_\tl \gamma_\mu \nu_\tl + \text{h.c.},
\end{eqnarray}
%%%%%%%%%%%%%%%%%%%%%%%%%%%%%%%%%%%%%%%%%%%%%%%%%%%%%%%%%%%%%%%%%%%%
gives rise to the $\nbb$-decay operator of \Fig{intrographs}a where
a light Majorana neutrino is exchanged.  In \Eq{SMLag},
$\gf=\sqrt{2}g^2/(8\mw^2)$ is the Fermi
constant, $\mw$ the charged weak 
boson mass, $g=0.65265$, $u$ and $d$ are the up and down quark fields
respectively, $e$ the  electron field and $\nu$ the neutrino field.
From the Feynman rules, the amplitude for this diagram is simply
%%%%%%%%%%%%%%%%%%%%%%%%%%%%%%%%%%%%%%%%%%%%%%%%%%%%%%%%%%%%%%%%%%%%
\begin{eqnarray}\label{lightx}
8 \gf^2\bar{u}_{\text{L}}\gamma^\mu d_\tl
\frac{\mn}{Q^2 -\mn^2}
\bar{u}_{\text{L}}\gamma_\mu d_\tl
\bar{e}^c_{\text{L}}e_\tl
~~~~~~~~~~~~~~~~~~~~& & 
\nonumber
\\
\cong
\frac{g^4}{16\mw^4}\frac{\mn}{Q^2}~
\bar{u}\gamma^\mu (1-\gamma^5) d~
\bar{u}\gamma_\mu (1-\gamma^5) d~
\bar{e}^c_\tl e_\tl
& &
\end{eqnarray}
%%%%%%%%%%%%%%%%%%%%%%%%%%%%%%%%%%%%%%%%%%%%%%%%%%%%%%%%%%%%%%%%%%%%
where $\mn$ is the neutrino mass.  

The current-current interaction of \Eq{SMLag} gives rise to
lepton-hadron vertices ($\pi \nu e$, $NN\nu e$) that
contribute to $\nbb$-decay through the operators shown in
\Fig{pionex}~\cite{Prezeau:2003xn}
%%%%%%%%%%%%%%%%%%%%%%%%%%%%%%%%%%%%%%%%%%%%%%%%%%%%%%%%%%%%%%%%%%%%
\begin{eqnarray}\label{hadronlepton}
{\cal{L}}_{h\nu e} \doteq 
\sqrt{2} \gf 
\left(
\fp \partial^\mu \pi^- ~\bar{e}_\tl \gamma_\mu \nu_\tl
\right.
~~~~~~~~~~~~~~~~~~~~& &
\nonumber
\\
\left.
+
\bar{p}\gamma^\mu (\gv - \ga\gamma^5) n~\bar{e}_\tl \gamma_\mu \nu_\tl
\right) 
+\text{h.c.} & &
\end{eqnarray}
%%%%%%%%%%%%%%%%%%%%%%%%%%%%%%%%%%%%%%%%%%%%%%%%%%%%%%%%%%%%%%%%%%%%
Operators that contribute to $\nbb$-decay are either suppressed or
enhanced by powers of $(p/\lh)^n$ (where $n$ is the {\it chiral power}
of the operator) with $p\!\sim$0.1~GeV and where
$\lh$ is a hadronic scale~$\sim$1~GeV.  The chiral power of a
$\nbb$-decay operator can be calculated with the following rules:
\begin{itemize}
\item  a derivative in a vertex counts as one power of $p$; 
\item pion and light neutrino propagators count as $p^{-2}$.
\end{itemize}
Considering that the parity-conserving pion-nucleon vertex is
$(\ga/\fp)\bar{N}\gamma^\mu\gamma^5 N\partial_\mu\pi$ and noting the
derivative 
in the pion-lepton operator in \Eq{hadronlepton},  one finds
that the $\nbb$-decay operators of \Fig{pionex} are all of
${\cal{O}}(p^{-2})$.  From Ref.~\cite{Prezeau:2003xn}, it is seen that
these operators have the same chiral power as the LO $\nbb$-decay
heavy particle exchange operators.  This observation suggests a more
precise method to compare heavy and light particle exchange
contributions to $\nbb$-decay.

Consider the amplitude for the Feynman graph of \Fig{pionex}a:
%%%%%%%%%%%%%%%%%%%%%%%%%%%%%%%%%%%%%%%%%%%%%%%%%%%%%%%%%%%%%%%%%%%%
\begin{eqnarray}\label{amplitude}
\text{\Fig{pionex}a}=~~~~~~~~~~~~~~~~~~~~~~~~~~~~~~~~~~~~~~~~~~~~~~~~~~~~
& &
\nonumber
\\
8\gf^2
\ga^2 M^2 \mn\frac{q_1\cdot q_2}{Q^2-\mn^2}
\frac{\bar{p}_1\gamma^5 n_3}{q_1^2-m_\pi^2}\frac{\bar{p}_2\gamma^5
  n_4}{q_2^2-m_\pi^2} 
\times
\bar{e}_\tl e^c_\tl
& & 
\nonumber
\\
\cong
8\gf^2
\ga^2 M^2 \mn
\frac{\bar{p}_1\gamma^5 n_3}{q_1^2-m_\pi^2}\frac{\bar{p}_2\gamma^5
  n_4}{q_2^2-m_\pi^2} 
\times
\bar{e}_\tl e^c_\tl,
~~~~~~~
& &
\end{eqnarray}
%%%%%%%%%%%%%%%%%%%%%%%%%%%%%%%%%%%%%%%%%%%%%%%%%%%%%%%%%%%%%%%%%%%%
where the error stemming from writing $q_1\cdot q_2/Q^2 \cong 1$ is of
${\cal{O}}(Q\cdot (k_1-k_2)/Q^2)$ with $|k_1+k_2|\cong
2.5$~MeV being the energy 
carried off by the electrons.  The approximation in
\Eq{amplitude} is therefore very good.  \Eq{amplitude} has the same
form as the LO $\nbb$-decay hadronic operators stemming from the exchange
of a heavy particle~\cite{Prezeau:2003xn}.  It follows that to a high
degree of precision, one can introduce a new ``short-distance''
$\nbb$-decay operator that stems from light neutrino exchange:
%%%%%%%%%%%%%%%%%%%%%%%%%%%%%%%%%%%%%%%%%%%%%%%%%%%%%%%%%%%%%%%%%%%%
\begin{eqnarray}\label{newpionop}
{\cal{L}}_{\pi\pi ee}  \doteq
2\mn\gf^2\fp^2\pi^-\pi^-\bar{e}_\tl e^c_\tl + \text{h.c.}
\end{eqnarray}
%%%%%%%%%%%%%%%%%%%%%%%%%%%%%%%%%%%%%%%%%%%%%%%%%%%%%%%%%%%%%%%%%%%%
This operator combined with the parity-conserving pion-nucleon vertex
yields \Eq{amplitude}.
In this form, comparing heavy and light particle exchange
contributions to $\nbb$-decay is relatively easy.  The only caveat
is the existence of a possible suppression of \Fig{pionex}a
relatively to \Fig{pionex}c due to the fact that the pion has a
finite range smaller than the size of the nucleus, while the
neutrino exchanged in the latter graph does not.  
In coordinate space,
the suppression of the nuclear matrix elements will occur through
exponentials of the form $e^{-m_\pi r}$.  In momentum space, the
suppression is due to a factors of the form $Q^2/m_\pi^2$.  The way to
handle this is discussed further below.

From Ref.~\cite{Prezeau:2003xn}, a LO operator has the
form
%%%%%%%%%%%%%%%%%%%%%%%%%%%%%%%%%%%%%%%%%%%%%%%%%%%%%%%%%%%%%%%%%%%%
\begin{eqnarray}\label{LOoperator}
\frac{\lambda^2}{\lbb^5} \bar{u}d~\bar{u}d~\bar{e}_\tl e^c_\tl
\end{eqnarray}
%%%%%%%%%%%%%%%%%%%%%%%%%%%%%%%%%%%%%%%%%%%%%%%%%%%%%%%%%%%%%%%%%%%%
where the general $\nbb$-decay vertex is assumed to be
suppressed by five powers of the $\nbb$-decay scale,
$\lbb$ as occurs in many popular particle physics
models.\footnote{Note that the authors of Ref~\cite{Cirigliano:2004tc} 
  insert an extra factor of $\gf^2\mw^4$ in \Eq{LOoperator}.}  For
example, this suppression by five powers occurs in R-parity violating
supersymmetry (RPV SUSY) and the left-right symmetric model
(LRSM).\footnote{In RPV SUSY, the coupling has the form
  $\lambda_{111}^{\prime 2}/\ls^5$ ($\lambda_{111}^{\prime }$ is the
  RPV SUSY coupling constant and $\ls$ is the SUSY breaking scale)
  while in the LRSM it has the form 
  $\zeta g^4/(32\mw^2 \mr^2 \nr)$ where $\mr$ is the mass of the
  right-handed boson, $\nr$ is the mass of the right-handed neutrino,
  and $\zeta<\mw^2/\mr^2$ is the mixing angle between $\wrt$ and $\wl$.}

This LO operator leads to the $\pi\pi ee$ operator
%%%%%%%%%%%%%%%%%%%%%%%%%%%%%%%%%%%%%%%%%%%%%%%%%%%%%%%%%%%%%%%%%%%%
\begin{eqnarray}\label{heavypionop}
\frac{\lambda^2}{\lbb^5} \bar{u}d~\bar{u}d~\bar{e}_\tl e^c_\tl
\to
\beta\frac{\lambda^2}{\lbb^5}\lh^2\fp^2\pi^-\pi^-\bar{e}_\tl e^c_\tl ,
\end{eqnarray}
%%%%%%%%%%%%%%%%%%%%%%%%%%%%%%%%%%%%%%%%%%%%%%%%%%%%%%%%%%%%%%%%%%%%
where $\beta$ is a parameter of ${\cal{O}}(1)$ generated by the
hadronization of the quark currents as discussed in
Ref.~\cite{Prezeau:2003xn}. 

Comparing \Eq{heavypionop} with \Eq{newpionop}, it is seen that the
contributions to $\nbb$-decay from light neutrino exchange and heavy
particle exchange processes are equal when
%%%%%%%%%%%%%%%%%%%%%%%%%%%%%%%%%%%%%%%%%%%%%%%%%%%%%%%%%%%%%%%%%%%%
\begin{eqnarray}\label{lightheavycomp}
\frac{\mn}{1~\text{eV}}=3.8\times 10^{3}
\frac{\lambda^2}{\alpha_{\text{M}}\lbb^5}~\text{TeV}^5 ,
\end{eqnarray}
%%%%%%%%%%%%%%%%%%%%%%%%%%%%%%%%%%%%%%%%%%%%%%%%%%%%%%%%%%%%%%%%%%%%
where $\alpha_{\text{M}}$ is a number that takes into account the fact
that the matrix elements of the pion exchange diagram of
\Fig{pionex}a is suppressed with respect to the matrix element of
the operator generated by the graph of \Fig{pionex}c.

In Ref.~\cite{Simkovic:1999re}, the nuclear matrix elements from
pseudoscalar couplings (corresponding to \Fig{pionex}a) and
axialvector coulings (corresponding to \Fig{pionex}c) were computed
for nine nuclei and tabulated in their Table~II.  On average, for
light neutrino exchange, the matrix elements stemming from
pseudoscalar coupling and denoted $M_{PP}^{\text{light}}$ in
Ref.~\cite{Simkovic:1999re} are ten times smaller than the matrix
elements stemming from axialvector coupling and denoted
$M_{AA}^{\text{light}}$.  The value $\alpha_{\text{M}}=10$ will
therefore be used.  Limits on $\lbb$ derived below are not very
sensitive to the exact value of $\alpha_{\text{M}}$ since $\lbb$
appears to the fifth power in \Eq{lightheavycomp}: factors of two or
three in $\alpha_{\text{M}}$ can change the results appearing below by
at most 1.2.

\begin{figure}[tb]
\resizebox{8 cm}{!}{\includegraphics{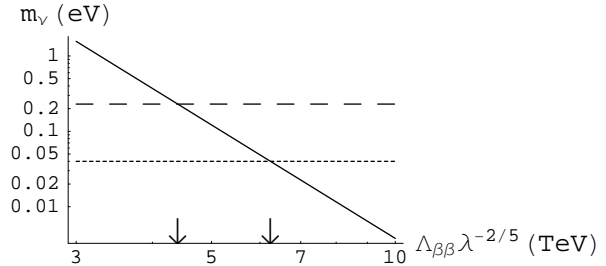}}
\caption{The solid line referred to in the text as the light-heavy
  equality (LHE) line, corresponds to the
  region in parameter space
  where light and heavy particle exchange contribute equally to the
  $\nbb$-decay amplitude.  The dashed line is the upper-limit on the
  neutrino mass from WMAP.  The dotted line represents the possible
  limit on the neutrino mass from the Planck mission in combination with
  the Sloan Digital Sky Survey (SDSS)~\cite{York:2000gk}.  The arrows
  are the values of 
  $\lbb\lambda^{-2/5}$ where the LHE line intersects
  the dashed and dotted lines.} \label{generalplot}
\end{figure}
%%%
%%%\begin{figure*}
%%%\begin{picture}(460,150)(0,0)
%%%\scalebox{0.8}[0.8]{\includegraphics{nonu_gen.eps}}
%%%\end{picture}
%%%\end{figure*}
%%%
\Eq{lightheavycomp} is plotted in \Fig{generalplot}.  Above the LHE
line, light neutrino exchange is larger than the heavy particle
contribution to $\nbb$-decay; the reverse is true below the LHE line.

The dashed line in \Fig{generalplot}
is the upper-limit on $\mn<$~0.23~eV from the WMAP~\cite{Spergel:2003cb}.
The upper-limit on  
  $\mn$ implies that the $\nbb$-decay heavy particle exchange operator
  is larger than the light neutrino exchange contribution for
%%%%%%%%%%%%%%%%%
\begin{eqnarray}\label{mod_ind_lim}
  \lbb\lambda^{-2/5}<4.4~\text{TeV},
\end{eqnarray}
%%%%%%%%%%%%%%%%
indicated by the first arrow in \Fig{generalplot}.  This is essentially a
model independent limit.  If
one uses the neutrino mass limit $\mn<0.04$~eV that could be reached by
the future Planck mission~\cite{Hannestad:2002cn}, and represented by
the dotted line in \Fig{generalplot}, then 
%%%%%%%%%%%%%%%%%
\begin{eqnarray}\label{planck_mod_ind_lim}
  \lbb\lambda^{-2/5}<6.2~\text{TeV},
\end{eqnarray}
%%%%%%%%%%%%%%%%
as indicated by the second arrow in the plot.

We can evaluate in specific models the point at which heavy particle
exchange becomes larger than light neutrino exchange.  Considering the
LRSM first, we
have~\cite{Mohapatra:1974gc,Senjanovic:1975rk,Prezeau:2003xn} 
%%%Taking
%%%$T^{\nbb}_{1/2}>10^{25}$~years 
%%%as a lower-limit on the $\nbb$-decay half-life we have in the LRSM:
%%%
%%%Since light neutrino exchange has not been ruled out as a possible
%%%mechanism for $\nbb$-decay, the region in parameter space where the
%%%heavy exchange contribution dominates should already be
%%%experimentally ruled out.  It follows that lower limits on
%%%$\lbb\lambda^{-2/5}$ derived from $\nbb$-decay in specific models
%%%should be greater or approximately equal to 4.4~TeV.  This can be
%%%checked in the LRSM and RPV-SUSY with $T^{\nbb}_{1/2}>10^{25}$~years
%%%as a lower-limit on the $\nbb$-decay half-life
%%%
%%%In the LRSM, for instance,
%%%%%%%%%%%%%%%%%%%%%%%%%%%%%%%%%%%%%%%%%%%%%%%%%%%%%%%%%%%%%%%%%%%%
\begin{eqnarray}
\frac{\lambda^2}{\lbb^5}&\equiv& \zeta\frac{g^4\mr^2}{32\mw^2}
\frac{1}{\mr^4\nr}
\\
\lambda^2 &\equiv& \zeta\frac{g^4\mr^2}{32\mw^2} < 5\times 10^{-4},
\label{smalllambda}
\end{eqnarray}
%%%%%%%%%%%%%%%%%%%%%%%%%%%%%%%%%%%%%%%%%%%%%%%%%%%%%%%%%%%%%%%%%%%%
where the limits $\zeta<10^{-3},~\mr>800$~GeV  on the weak gauge boson
mixing angle and the right-handed weak boson mass were used in
evaluating \Eq{smalllambda}.  Thus, the
upper-limit on the right-handed particle masses below which the heavy
particle exchange contribution dominates is  $\mr\sim\nr\sim \lr <
0.9$~TeV.  Using the lower-limit on the half-life of $\nbb$-decay of
$T^{\nbb}_{1/2}>10^{25}$~years, one obtains a lower-bound on the mass 
of the heavy right-handed particles of $\sim
1$~TeV~\cite{Prezeau:2003xn}; this implies that the region in
the LRSM parameter space where the heavy particles dominate over the
light neutrino exchange contributions is essentially ruled out for the
current limits on $\mn$.

Similarly in RPV-SUSY, the diagram that provides the strongest
constraints on $\lambda_{111}^\prime$ is the one with gluino
exchange shown in \Fig{intrographs}b.  From
Ref.~\cite{Faessler:1996ph}, we have
%%%%%%%%%%%%%%%%%%%%%%%%%%%%%%%%%%%%%%%%%%%%%%%%%%%%%%%%%%%%%%%%%%%%
\begin{eqnarray}
\frac{\lambda^2}{\lbb^5} \equiv
\alpha_{\text{S}}\frac{8\pi}{9}\frac{\lambda_{111}^{\prime
    2}}{m_{\tilde{q}}^4 m_{\tilde{g}} }. 
\end{eqnarray}
%%%%%%%%%%%%%%%%%%%%%%%%%%%%%%%%%%%%%%%%%%%%%%%%%%%%%%%%%%%%%%%%%%%%
Taking $m_{\tilde{q}}= m_{\tilde{g}}=\ls=1$~TeV,
$\alpha_{\text{S}}=1$, and substituting in \Eq{mod_ind_lim}, one
obtains that the heavy particle exchange contribution is largest when
$\lambda_{111}^{\prime }> 1.5\times 10^{-2}$.
The lower-limit on $\lambda_{111}^{\prime }$ extracted from
$\nbb$-decay and $T^{\nbb}_{1/2}>10^{25}$~years 
is $\lambda_{111}^{\prime }< 6.3\times 10^{-2}$ and
$\lambda_{111}^{\prime }< 10^{-2}$ derived from $\nbb$-decay in
Ref.~\cite{Faessler:1996ph} and Ref.~\cite{Wodecki:1998vc}
respectively.  Hence, the region in RPV-SUSY where the heavy particle
exchange contribution dominates over the light neutrino exchange is
also ruled out for current limits on $\mn$.

Using the limit in \Eq{planck_mod_ind_lim} instead, one obtains
%%%%%%%%%%%%%%%%%%%%%%%%%%%%%%%%%%%%%%%%%%%%%%%%%%%%%%%%%%%%%%%%%%%%
\begin{eqnarray}
\lr < 1.4~\text{TeV},~~~
\lambda_{111}^{\prime }> 6.3\times10^{-3}
\end{eqnarray}
%%%%%%%%%%%%%%%%%%%%%%%%%%%%%%%%%%%%%%%%%%%%%%%%%%%%%%%%%%%%%%%%%%%%
instead.

Note that with current limits on $\nbb$-decay, one requires the
coupling constant $\lambda$ in \Eq{LOoperator} to be $\ll 1$ if one
demands that $\lbb \cong 1$~TeV.  This observation is here verified in
the LRSM and RPV SUSY.  Although such a small value of $\lambda$ is
clearly allowed, it should be explained since naturalness suggests
$\lambda \sim 1$ instead.  In this case, one would expect $\lbb >
4.4$~TeV.

In this note, a simple and precise method of comparing contributions to
$\nbb$-decay was presented.  It was shown that LO $\pi\pi e^-e^-$
operators can be written down for both light neutrino exchange and
heavy particle exchange contributions to $\nbb$decay.  This observation
facilitated their comparison and allowed us to plot a graph in the
neutrino mass and heavy particle scale $\lbb$ parameter space to discern
the regions where one contribution may be
larger than the other.  Using current limits on $\nbb$-decay, it was
also shown that $\lbb\gtrsim 4.4$~TeV in a general
particle physics model 
where the $\nbb$-decay operator coupling constant is assumed to be of
${\cal{O}}(1)$.

The author would like to thank Petr Vogel and Vincenzo Cirigliano for
useful discussions and comments in the preparation of this manuscript.

%%%%%%%%%%%%%%%%%%%%%%%%%%%%%%%%%%%%%%%%%%%%%%%%%%%%%%%%%%%%%%%%%%%%

\end{document}